# Astro2020 Science White Paper

# HIGH DEFINITION ASTROMETRY

**Thematic Areas:**  ☒ Planetary Systems   ☒ Star and Planet Formation

**Principal Author:**
Name:   Philip Horzempa
Institution:  LeMoyne College
Email:  horzempa45@gmail.com
Phone:  605-481-8688

**Abstract**:

   High Definition Astrometry (0.1 - 1.0 micro-arcseconds) will open a new window into neighboring planetary systems.  For the first time, the realm of temperate terrestrial worlds will be explored.  This includes Earth Analogs, thereby allowing the value of eta-Earth to be directly determined, without resort to extrapolation.

   High Definition Astrometry will provide a means to confirm the existence of Radial Velocity (RV) planets while, at the same time, measuring true mass, along with orbit inclination and radius, i.e., system architecture.  Planetary systems not amenable to RV search, such as those in a "face-on" orientation, will be surveyed for the first time.

   Extreme Precision Astrometry is not only useful, but is essential to the future of exoplanet research.

**HIGH DEFINITION ASTROMETRY**

With the release of Gaia's DR2 on April 25, 2018, the field of exoplanet research has entered the era of High Definition astrometry. The first era of Radial Velocity (RV) and Transit measurements has revealed the variety of planets and the rich diversity of alien solar system architectures.  Astrometry will compliment RV and Transit discoveries by expanding their discovery space.

The stage is now set for the next advance, an era of very precise astrometry, reaching levels as fine as 0.1 micro-arcsecond (uas) [100 nano-arcsecond (nas)], astrometry.   That is a precision 50-100 times better than that of Gaia and it will open up the realm of temperate terrestrial worlds.

The hardware for a 100-nas mission was designed, constructed and validated under the auspices of NASA's SIM (Space Interferometry Mission) project.  By 2010, after an investment equal to "one Kepler telescope" ($600 million), SIM had completed Phase A and Phase B, and achieved TRL-8 (**1**).  A new Probe-class mission, Star Watch (P. Horzempa, in preparation), will build on this remarkable national technical heritage, incorporating advances in laser metrology engineering since 2010.

**Overview**

Contributions of high-definition astrometry:

\>\>  Exoplanet system architecture completeness

\>\>  Detection of small planets, including Earth Analogs

\>\>  Confirmation of RV planets

\>\>  True mass of RV worlds

Earlier publications have documented the reach of this level of astrometry, for both exoplanet research and astrophysics in general **(2)**.  This essay will consider examples of how high-definition astrometry will address specific issues and stellar targets.

**Gaia Early Observations**

It is instructive to consider how Gaia has already contributed to a more complete picture of several nearby exoplanet systems. **(3).**

For Barnard's Star, tau Ceti and Ross 128, Gaia detected a hints of velocity anomalies which could be explained by Giant planets in distant orbits.  For Epsilon Indi, a significant velocity acceleration is observe by

Gaia, and is compatible with the recently announced long-period gas giant planet orbiting this star.  For 51 Pegasi, a strong velocity acceleration was observed, indicating the presence of a 6 Jupiter-mass planet in 3 AU orbit, up to a brown dwarf of several tens of Jupiter-masses in a >100 AU orbit could explain this.  This system, famous for the first Hot Jupiter, is now shown to have a more complex geometry.

Lastly, they detect the astrometric signal of the known planet orbiting Beta Pictoris.

In one fell swoop, Gaia has expanded our view of these systems demonstrating the reach of Astrometry.  It is able to probe known planetary systems and reveal what has been previously hidden.

### Relevance to NAS Exoplanet Report (4)

As with Gaia, the 100-nas astrometry probe's parameter space will overlap significantly with that of radial velocity surveys, and will inform us of the masses and the relative inclinations of the orbits in multiplanet systems. It will detect planets orbiting stars more massive than those probed by radial velocity surveys.

The key "skill" of Star Watch is that it will provide access to the population of planets that are of either lower mass or are more distant than Jupiter, particularly for Sun-like stars. Star Watch will probe the majority of the parameter space that is spanned by the Solar System planets, and will be sensitive to worlds that have not already been surveyed by radial velocity surveys.  To quote the NAS report: "An understanding of planet formation is critically informed by the existence of planets with masses less than, and separations greater than, Jupiter."

The NAS report listed remaining Science Gaps in the Planetary Census and pointed out that the "demographics of planets orbiting young stars would be exceptionally valuable" allowing insight into formation processes. The activity of young stars makes planet detection difficult for most methods but has essentially no effect on astrometry allowing it to detect planets previously hidden

### The G-Dwarf Case

The NAS report stated that there is a compelling need to develop the means to detect and characterize temperate terrestrial planets orbiting Sun-like stars.  High-Definition Astrometry is the quickest path to achieving that goal and will alleviate the costly requirement, in terms of mission time, for future direct imaging missions to self-discover their own targets.

Detecting an Earth-size planet in the habitable zone of a Sun-like star at a distance of 10 pc is within the reach of a 100-nas astrometry probe.  As the NAS report pointed out, such detection is "essentially impossible with any other technique" and "it is extremely unlikely that such a planet will be found to transit any nearby star."

The NAS report points out that the discovery of dozens of terrestrial planets will reveal the diversity of this class of planet. The Star Watch High-Definition Astrometry mission would make this possible

**Ambiguous RV Systems**

Many planetary systems detected by Doppler teams are not well defined. This is a result of only minimum mass determination for those planets. A common error is to assume that m(sin i) is the true mass which can lead to misconceptions as to their true nature.

One example of an ambiguous system is tau Ceti, an older twin of the Sun. It has been referred to as a system of 4 super Earths. **(5)** This would be the case only if the RV m(sin i) mass values are the true values. However, if one assumes that the invariant plane for the system is the same as that of the observed disk (40 degrees), then the masses would range from 2.5 to 10 Me, putting them in the sub-Neptune to Neptune class. That would still be an interesting system but it would not be a system of super Earths. High-Definition astrometry will bring clarity, as these inner worlds of tau Ceti, if they exist, would generate signals that ranged from 1 to 15 uas, well within the reach of the next-generation 100 nano-arcsecond Probe. As a bonus, the Star Watch mission would be able to detect any true terrestrial worlds in the tau Ceti system.

**Verification of RV Planets**

Confusion about the true architecture of planetary systems can result from the challenging task of teasing signals from Radial Velocity data. The existence of the planets themselves is, at times, in question. High Precision Astrometry would provide the crucial aspect of confirmation, something that is missing for many RV planets.

Let's again consider the tau Ceti system. It was first reported that the star was orbited by four super Earths, but a later RV survey could only detect reliable signs of two of those worlds. **(5)** That new survey, however, found two planets not detected by the first survey. This has led to confusion as to whether there are 2 or 4 or 6 planets orbiting the star.

Another example is the collection of planets orbiting HR 8832. It is an early K star located 7 parsecs away. A total of four independent Radial Velocity studies regarding the planetary system of HR 8832 have been done, with many of their results conflicting with each other. A total of 7 planet detections have been claimed, but 2 are thought to be spurious.

A recent paper reviewed RV data for several single planets in eccentric orbits. **(6)** To quote the paper: "Of particular interest is the question of the multiplicity in planetary systems discovered using sparse RV data. With limited data, it can be possible to miss the presence of additional planets, by focusing too heavily on a single, dominant signal. At the same time, it is possible to find 'planets' that do not actually exist, by identifying periodicities in the data that either vanish with the acquisition of more observations, or turn out to be the result of other astrophysical phenomena."

Further: "Science is the pursuit of truth. In that spirit, it remains critically important to revise our understanding of the architectures of planetary systems as new data become available. There is a vast body of literature concerning the analysis of RV data for planetary signals in the midst of the confounding effects of sparse sampling and stellar noise. Of particular relevance is work that has re-analysed data on known planetary systems to confirm, clarify, or refute the published planetary parameters." High Definition Astrometry would provide data to confirm the existence of RV planets and determine the true configuration of RV exoplanetary systems.

**Protoplanetary Disks**

Recent images of concentric and spiral structure in protoplanetary disks show patterns that may be due to the presence of planets, both large and small. This is a field in which astrometry can provide unique data. Since an astrometric probe observes only the motion of the parent star, its measurements are unaffected by the stellar disk. It could detect embedded worlds and determine their mass and orbits.

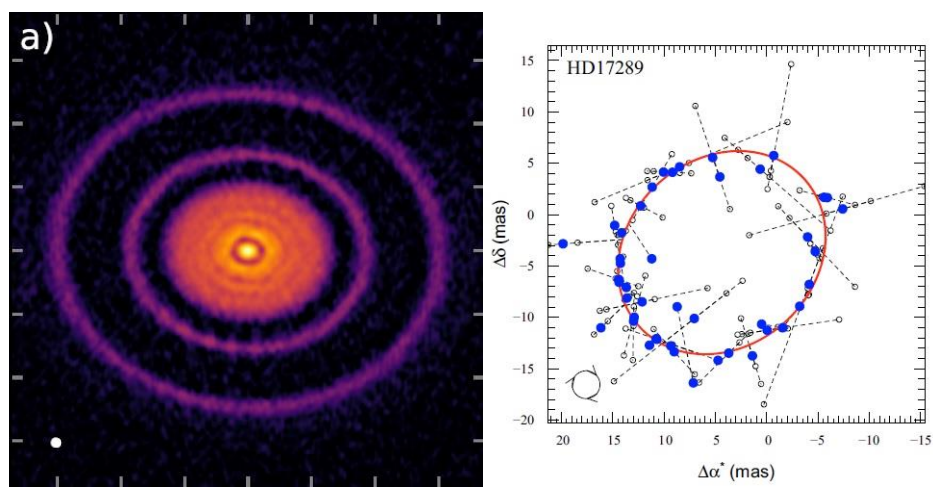

Figure 1: a.) AS 209 disk; b.) astrometric solution for Brown Dwarf

**"Empty" and Face-On Star Systems**

Several nearby Solar Analogs appear to be planetless, as observed by transit and RV methods. It is not known if the absence of planets is an accurate picture or if their worlds are too small, or in orbits too inclined, to be detected. One example is zeta Reticuli, a neighbor only 20 light-years (6 parsecs) away. This is a very wide binary, consisting of G4 and G5 stars. At that distance, a 100-nas astrometry Probe would readily detect Earth Analogs orbiting each star. Another example of a "barren" star is phi$^2$ Pavonis, a single Solar Twin located only 7 parsecs away.

These may be examples of "face-on" systems. If so, then the high inclinations of planet orbits would make detection of their worlds very challenging, if not impossible, for RV efforts. Astrometry is essentially the only detection method by which we can determine if these "barren" systems really do harbor a collection of planets.

Another example of a "face-on" system of interest is the HR 8799 system. High precision astrometry can detect Super Earths lying within the orbit of the innermost Giant planet, providing a more complete picture of that system's architecture.

**Timing**

One of the advantages of flying a High-Definition Astrometry mission in the 2020s is timeliness. Schedule and cost issues of recent space projects, combined with attempts to cancel WFIRST, indicate that it is likely that 25 years will elapse before the launch of a Flagship space telescope. One key project of such a telescope is the search Earth Analogs in nearby star systems. Star Watch will allow that search to take place in the 2020s.

**Conclusions**

A significant Finding of the NAS report was that "High-precision, narrow-angle astrometry could play a role in the identification and mass measurement of Earth-like planets around Sun-like stars, particularly if the radial velocity technique is ultimately limited by stellar variability." **(4)**

Astrometry is the "ideal" method of detecting planetary systems, in that is sensitive to planets regardless of their inclination. This allows many, or even most, of the worlds in a system to be "cataloged," depending on the precision limits. A 100-nas mission opens up the realm of truly terrestrial worlds for the first time. Earth, Venus and Mars Analogs, in a G-star's Habitable Zone, will be detectable. The Star Watch astrometry probe will reveal new knowledge about the demographics of exoplanets. It will go a long way towards answering the question: What are the different kinds of planetary systems, and what are their relative frequencies of occurrence?

It will measure the true mass of the planets it detects (not just the minimum mass measured by radial velocity), it will provide knowledge of the coplanarity of exoplanet systems, and it will discover extremely rare (and valuable) planetary system "oddballs."